\documentclass[a4paper,twocolumn,superscriptaddress,prd]{revtex4}
\usepackage{epsf} 
\usepackage{epsfig}
\usepackage{latexsym}
\usepackage{amsfonts}
\usepackage{dcolumn}
\usepackage{varioref}
\usepackage{ifthen}
\usepackage{graphicx}
\usepackage{amssymb,amsmath,amsthm}
\usepackage{graphicx}
\usepackage{mathrsfs}
\usepackage{caption}
\usepackage{slashed}            
\usepackage{bbm}                
\usepackage{enumitem}
\usepackage{braket} 
\usepackage{float}

\newcommand{\INTa}{\int \dfrac{d^nk}{\left(2\pi\right)^n}}
\newcommand{\da}{\left(k^2-m^2 \right)}
\newcommand{\db}{\left((k+p)^2-m^2 \right)}

\newcommand{\mn}{^{\mu\nu}}
\newcommand{\meask}{\mathfrak{D}^nk}
\newcommand{\measkq}{\mathfrak{D}^nk\mathfrak{D}^nq}
\newcommand{\intkq}{\int\mathfrak{D}^nk\mathfrak{D}^nq}
\newcommand{\intk}{\int\mathfrak{D}^nk}
\newcommand{\transv}{\left(g^{\mu\nu}-\dfrac{p^{\mu}p^{\nu}}{p^2}\right)}

\newcommand\this{\addtocounter{equation}{1}\tag{\theequation}}
\newcommand{\mpi}{m^2} 

\newcommand{\eq}[1]{\begin{align*}#1 \end{align*} }
\newcommand{\li}{\text{Li}_2}
\newcommand{\m}{m^2}
\newcommand{\M}{M^2}
\newcommand{\eqn}[1]{\begin{align} #1 \end{align}}
\newcommand{\p}{p^2}
\newcommand{\ra}{r_1}
\newcommand{\rb}{r_2}
\newcommand{\rta}{\tilde{r}_{1}}
\newcommand{\rtb}{\tilde{r}_{2}}
\begin{document}
\parindent 0mm 
\setlength{\parskip}{\baselineskip} 
\thispagestyle{empty}
\pagenumbering{arabic} 
\setcounter{page}{1}
\mbox{ }
\title{Kroll-Lee-Zumino quantum field theory of pionic interactions: rho-meson  propagator at two-loop level}

\author{C. A. Dominguez}
\affiliation{Centre for Theoretical \& Mathematical Physics, and Department of Physics, University of
Cape Town, Rondebosch 7700, South Africa}
\author{M. Lushozi}
\affiliation{Centre for Theoretical \& Mathematical Physics, and Department of Physics, University of
Cape Town, Rondebosch 7700, South Africa}
\author{K. Schilcher}
\affiliation{Centre for Theoretical \& Mathematical Physics, and Department of Physics, University of
Cape Town, Rondebosch 7700, South Africa}
\affiliation{PRISMA Cluster of Excellence, 
Institut f\"{u}r Physik,
Johannes Gutenberg-Universit\"{a}t, 
D-55099 Mainz, Germany}
\affiliation{National Insitute for Theoretical Physics, 
Private Bag X1, Matieland 07602, South Africa}
\begin{abstract}
\noindent	
The Kroll-Lee-Zumino renormalizable Abelian quantum field theory of pionic strong interactions is used to compute the rho-meson propagator at the two-loop level. 
\end{abstract}
\maketitle
\noindent
\section{Introduction}
A renormalizable, Abelian, quantum field theory of strong interactions among pions and a massive neutral rho-meson was proposed long ago by Kroll, Lee, and Zumino (KLZ) \cite{KLZ1}. In spite of the presence of a massive gauge boson this KLZ theory is renormalizable, due to this boson coupling to a conserved current \cite{Hees}. An attractive feature of the theory is that it provides the quantum field theory justification for the Vector Meson Dominance (VMD) model \cite{VMD}. In addition, it is a potential candidate to fill the energy gap between chiral perturbation theory at threshold and QCD above $1 \, {\mbox{GeV}}$. A successful application was made some time later with the calculation of the rho-meson self energy at the one-loop level \cite{GK}. In fact, after using this result in the VMD expression of the electromagnetic pion form factor, a good agreement was found with data in the time-like region. In particular, this form factor agrees with the well known  Gounaris-Sakurai formula \cite{GS}-\cite{tau} in the vicinity of the rho-meson peak. This agreement is intriguing, given the fact that this formula is purely empiric. Another successful application is that of the pion form factor in the space-like region, determined from the triangle diagram \cite{CAD1}. There is excellent agreement with data up to $q^2 \simeq - \, 10 \,{\mbox{GeV}^2}$, with a chi-squared per degree of freedom of $\chi^2|_{\mbox{\small{KLZ}}} \, = 1.1$, in contrast with the VMD value of $\chi^2|_{\mbox{\small{VMD}}} \, = 5.0$. Furthermore, the pion mean-squared radius is predicted to be $\langle r^2_\pi \rangle|_{\mbox{\small{KLZ}}} = 0.46\, {\mbox{fm}}^2$, compared with the experimental value \cite{PDG}-\cite{rpiEXP} $\langle r^2_\pi \rangle|_{\mbox{\small{EXP}}} = 0.439\,\pm\, 0.008 \, {\mbox{fm}}^2$, and the VMD prediction $\langle r^2_\pi \rangle|_{\mbox{\small{VMD}}} = 0.39\, {\mbox{fm}}^2$. Finally, two equally successful applications of KLZ are
the scalar radius of the pion \cite{CAD2}, and the scalar form factor of the pion in the space-like region \cite{CAD3}, both in good agreement with Lattice QCD \cite{LQCD1}-\cite{LQCD2}.\\
Motivated by these results we calculate in this paper the KLZ rho-meson propagator at the two-loop level in perturbation theory.\\

The KLZ Lagrangian is given by
\begin{eqnarray}
\mathcal{L}_{KLZ} &=& \partial_\mu \phi \, \partial^\mu \phi^* -  m^2 \,\phi \,\phi^* - \tfrac{1}{4}\, \rho_{\mu\nu} \,\rho^{\mu\nu} 
+ \tfrac{1}{2}\, M^2\, \rho_\mu \,\rho^\mu \nonumber\\ [.3cm]
&+&g_{\rho\pi\pi} \rho_\mu J^\mu_\pi\ + g_{\rho\pi\pi}^2 \;\rho_\mu\; \rho^\mu \;\phi \;\phi^*   \;,
\end{eqnarray}
with $m$ and $M$ the pion and rho-meson masses, respectively,
$\rho_\mu$  a vector field of the $\rho^0$ meson ($\partial_\mu \rho^\mu = 0$), $\phi$  a complex pseudo-scalar field describing the $\pi^\pm$ mesons, $\rho_{\mu\nu}$ is the  field strength tensor, $\rho_{\mu\nu}  = \partial_\mu \rho_\nu - \partial_\nu \rho_\mu$, and $J^\mu_\pi$ is the $\pi^\pm$ current, $J^\mu_\pi  = i \phi^{*} \overleftrightarrow{\partial_\mu} \phi$.
In spite of the explicit presence of the $\rho^0$ mass term above, the theory is renormalizable. This is due to the neutral vector meson being coupled to a conserved current \cite{KLZ1}-\cite{Hees}.\\
The $\rho$-meson self energy at the one-loop level, Fig.1, is given by \cite{GK}
\begin{align} 
i \,\Pi^{\mu\nu}(p^2)&=g^2 \int\frac{d^4k}{(2\pi)^4}\left\lbrace\frac{(2k+p)^{\mu}(2k+p)^{\nu}}{(k^2-m^2)((k+p)^2-m^2)}  \right.  \nonumber\\
&\left. - 2\,  \frac{g^{\mu\nu}}{k^2-m^2}\right\rbrace =\left(g^{\mu\nu}-\frac{p^{\mu}p^{\nu}}{p^2} \right)i \, F_{\text{vac}}(p^2) \;,
\end{align}	
where the vacuum polarization, $F_{\text{vac}}(p^2)$, in the time-like region, after dimensional regularization and renormalization, is given by \cite{GK}
\eq{&F_{\text{vac}}(p^2)=\frac{g^2 \, p^2}{48 \,\pi^2}\Bigg\lbrace \left(1-\frac{4\m}{\p}\right)^{\frac{3}{2}} \Bigg[\ln \left|\frac{1+\sqrt{1-\frac{4\m}{\p}}}{1-\sqrt{1-\frac{4\m}{\p}} } \right|\\
      &-i\pi\theta\left( \p-4\m \right) \Bigg]+\frac{8\m}{\p}+\left(\frac{4\m}{\p}-1\right)^{\frac{3}{2}} \\
      & \times\cos^{-1}\left(1-\frac{\p}{2\m}\right)\bigg[\theta(\p-4\m) -\theta(\p)\bigg]+C\Bigg\rbrace\,,  \this \label{galekapustaFormula}
      }
      where
      \eq{C&=-\frac{8\m}{\M}+\left(1-\frac{4\m}{\M}\right)^{3/2}\ln\left|\frac{1-\sqrt{1-\frac{4\m}{\M}}}{1+\sqrt{1-\frac{4\m}{\M}}}\right|\,\, .\this  
      } 

\begin{figure}[ht]
\begin{center}
\includegraphics[width=3.3in, height=3.9in]{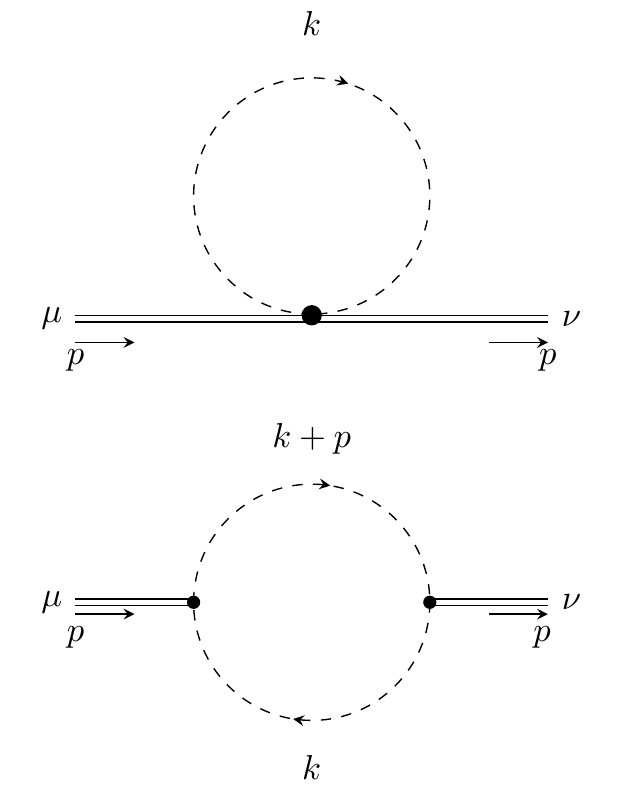}
\caption{Rho-meson vacuum polarization. Solid/broken lines correspond to the rho-meson/pion.}
\label{figure1}
\end{center}
\end{figure}
This vacuum polarization function is related to experimental data on electron-positron annihilation into hadrons as follows. The electromagnetic correlation function is defined as
\begin{eqnarray}
\Pi^{\mu\nu}(s) &=& i \int d^4x \, e^{-i p \cdot x} \, \langle 0| T(J^{\mu}_{\mbox{\footnotesize{EM}}}(x) J^{\nu}_{\mbox{\footnotesize{EM}}}(0)|0\rangle \nonumber \\ [.3cm]
&=& (- g^{\mu \nu}\; s + p^{\mu} p^{\nu}) \;  \Pi_{\mbox{\footnotesize{EM}}}(p^2 \equiv s)\, ,
\label{key}\end{eqnarray}
where $J^{\mu}_{\mbox{\footnotesize{EM}}}(x)$ is the vector-isovector current. Invoking the current-field identity
\begin{equation}
J^{\mu}_{\mbox{\footnotesize{EM}}}(x)= - \frac{M}{f_\rho} \rho^{\mu}(x) ,
\end{equation}
where $f_\rho = 4.97 \pm 0.07$  \cite{PDG}, the function $\Pi_{\mbox{\footnotesize{EM}}}(s)$ in terms of the vacuum  polarization becomes
\begin{eqnarray}
\Pi_{\mbox{\footnotesize{EM}}}(s)&=& \frac{M^4}{s f_\rho^2} \frac{1}{[M^2 - s + \mbox{Re}\, F_{\mbox{vac}}(s)]^2 +[\mbox{Im} \,F_{\mbox{vac}}(s)]^2} \nonumber \\ [.3cm]
& \times& [M^2 - s + {\mbox{Re}} \,F_{\mbox{vac}}(s) - i\, \mbox{Im}\, F_{\mbox{vac}}(s)]\,.
\end{eqnarray}
Alternatively, the electromagnetic form factor of the pion in the framework of VMD can be expressed in terms of the vacuum polarization as
\begin{equation}
F_{\pi}(s)= \frac{M^2 +F_{\mbox{vac}}(0)}{M^2 - s + F_{\mbox{vac}}(s)}\,. \label{PionFF}
\end{equation}
This pion form factor, with $F_{\mbox{vac}}(s)$  computed in KLZ at the one-loop level \cite{GK}, agrees reasonably well with data in the time-like region. In fact, this agreement is better than that from plain (tree-level) VMD, i.e. for $F_{\mbox{vac}}(p^2)=0$. This, together with the agreement in the space-like region mentioned earlier, provides enough motivation to proceed to the two-loop level in the sequel.
\section{KLZ Lagrangian and Feynman Rules}	
\begin{figure*}[t!]
\centering
\includegraphics{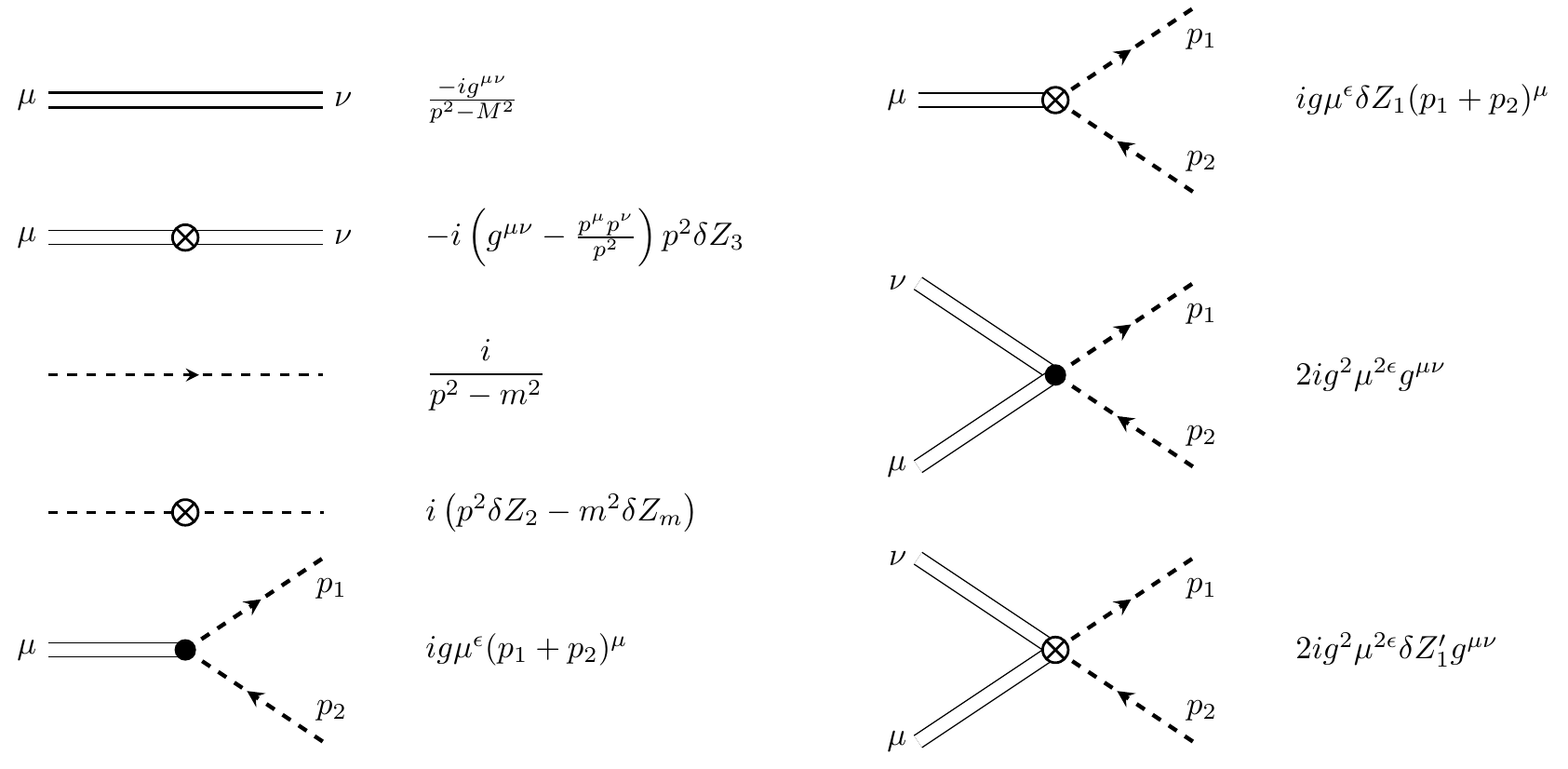}
\caption{\label{feynrules1} Full set of KLZ Feynman rules in the Feynman gauge. } 
\end{figure*}	
The gauge invariance of the KLZ theory can be revealed using St\"uckelberg's procedure \cite{stueckelberg1938forces1,Hees}, to wit. Quantizing KLZ as a gauge theory in the path-integral formalism introduces a gauge-fixing parameter, $\xi$, so that the Lagrangian to be used in pertubative calculations is 
\eq{\mathcal{L}_{\textbf{KLZ}}&=\partial_{\mu}{{\phi} _0^{\ast}}\, \partial^{\mu}{{\phi} _0} - m_0^2 \,{{\phi} _0^{\ast}} \,{\phi} _0 - \frac{1}{4} \, {\rho_0}^{\mu\nu}{\rho_0}_{\mu\nu} \\&
+\frac{1}{2}M_{0}^2 \, {\rho_0}^{\mu}{\rho_0}_{\mu}	+i \, g_0 \, \mu^\epsilon {\rho_0}^\mu \left({{\phi} _0}^\ast \overleftrightarrow{\partial_\mu} {\phi} _0\right)\\ &
+g_0^2 \, \mu^{2\epsilon} \,{\rho_0}^{\mu}{\rho_0}_\mu \, {{\phi} _0}^\ast{\phi} _0-\frac{1}{2\xi_0}\left(\partial_\mu {\rho_0}^\mu\right)^2\,\, .\this\label{LKLZ}
	}
Notice the appearance of the renormalization scale $\mu$, due to the Lagrangian being written in $n=4-2\,\epsilon$ dimensions. Each of the unrenormalized fields, $\phi_0$ and $\rho_0^{\mu}$,  has a zero-subscript to distinguish it from their renormalized counterparts, $\phi$ and $\rho^{\mu}$, respectively. Similarly, the bare parameters $m_0,M_0,g_0$ and $\xi_0$ are written with a zero-subscript whereas their physical/renormalized counterparts,  $m,M,g$ and $\xi$, are written without subscripts. To relate bare quantities to their renormalized partners we introduce renormalization constants as follows
\eq{\phi_0&=\sqrt{Z_2} \,\phi\,\, ,&  \rho^{\mu}_0&=\sqrt{Z_3} \, \rho^{\mu}\,\, ,
&Z_1 g&=g_0 \,Z_2\sqrt{Z_3} \,\, ,}
\eq{Z'_1 \,g^2&=g^2_0\, Z_2\, Z_3\,\, ,& Z_m \,m^2 &= m_0^2 \,Z_2\,\,, & Z_M \,M^2&= M_0^2\, Z_3 \,\, ,\\
\frac{1}{\xi}Z_{\xi}&=\frac{1}{\xi_0}Z_3\,\, , & & & & \this \label{ren-constants}	
} 
so that the Lagrangian, Eq.\eqref{LKLZ}, expressed in terms of renormalized quantities becomes
\eq{\mathcal{L}_{\textbf{KLZ}}&=Z_2\partial_{\mu}{\phi}^{\ast}\partial^{\mu}{\phi} -Z_mm^2{\phi}^{\ast}\phi - \frac{1}{4}Z_3{\rho}^{\mu\nu}{\rho}_{\mu\nu} \\&
+\frac{1}{2}Z_MM^{2}{\rho}^{\mu}{\rho}_{\mu}	+iZ_1g\mu^\epsilon {\rho}^\mu \left({\phi}^\ast \overleftrightarrow{\partial_\mu} \phi\right)\\ &
+Z_1'g^2\mu^{2\epsilon}{\rho}^{\mu}{\rho}_\mu{\phi}^\ast\phi-\frac{1}{2\xi}Z_{\xi}\left(\partial_\mu {\rho}^\mu\right)^2\,\, .\this\label{LKLZ1}
}
Each renormalization constant can be written as the unity plus a counter-term
\eq{Z_1&=1+\delta Z_1,& Z_1'&=1+\delta Z_1',& Z_2&=1+\delta Z_2 \this\label{deltaZ2}\\
Z_3&=1+\delta Z_3\,,& Z_M&=1+\delta Z_M\,,& Z_{\xi}&=1+\delta Z_{\xi}\this\label{deltaZ3}\\
Z_{m}&=1+\delta Z_m \,\, .	&&&&\this}
 Each counter-term is expressed as an expansion in  the coupling constant, $g$, i.e.
 \eqn{\delta Z_2&=\sum_{k=1}^{\infty}\delta Z^{(k)}_{2}\left(\frac{g^2}{16\pi^2} \right)^k \,\, ,\this\label{Z2series}\\
 	 \delta Z_3&=\sum_{k=1}^{\infty}\delta Z^{(k)}_{3}\left(\frac{g^2}{16\pi^2} \right)^k\,\, ,\this\label{Z3series}\\
 	 \delta Z_m&=\sum_{k=1}^{\infty}\delta Z^{(k)}_{m}\left(\frac{g^2}{16\pi^2} \right)^k\,\,,\this \label{Zmseries}
 }
where each coefficient of the expansion is to be determined order-by-order in perturbation theory.\\
The KLZ theory, being gauge invariant, has associated Ward identities which can be used to relate the counter-terms as
\eqn{\delta Z_1=\delta Z_2=\delta Z_1' \,\, ,\\
\delta Z_M =\,\,0\,\,= \delta Z_{\xi}\,\, .
}
This greatly simplifies the lagrangian and therfore all calculations. As a further simplification we make the gauge choice $\xi=1$, the so-called Feynman gauge, which leads to
\eq{ &\mathcal{L}_{KLZ}=\partial_{\mu}{\phi}^{\ast}\partial^{\mu}{\phi} -m^2{\phi}^{\ast}\phi - \frac{1}{4}{\rho}^{\mu\nu}{\rho}_{\mu\nu} +\frac{1}{2}M^{2}{\rho}^{\mu}{\rho}_{\mu}\\
	&+ig\mu^\epsilon {\rho}^\mu \left({\phi}^\ast \overleftrightarrow{\partial_\mu} \phi\right)
	+g^2\mu^{2\epsilon}{\rho}^{\mu}{\rho}_\mu{\phi}^\ast\phi-\frac{1}{2}\left(\partial_\mu {\rho}^\mu\right)^2\\
	&+\delta Z_2\partial_{\mu}{\phi}^{\ast}\partial^{\mu}{\phi} -m^2\delta Z_m{\phi}^{\ast}\phi - \frac{1}{4}\delta Z_3{\rho}^{\mu\nu}{\rho}_{\mu\nu} \\
	&+ig\delta Z_2\mu^\epsilon {\rho}^\mu \left({\phi}^\ast \overleftrightarrow{\partial_\mu} \phi\right)+g^2\delta{Z_2}\mu^{2\epsilon}{\rho}^{\mu}{\rho}_\mu{\phi}^\ast\phi \,
	.\this\label{LKLZFinal}
}
The Feynman rules that follow from Eq.\eqref{LKLZFinal} are shown in Figure \ref{feynrules1}.
\section{Self-energy of the rho-meson} 
\begin{figure*}[t]
	\centering
	\includegraphics{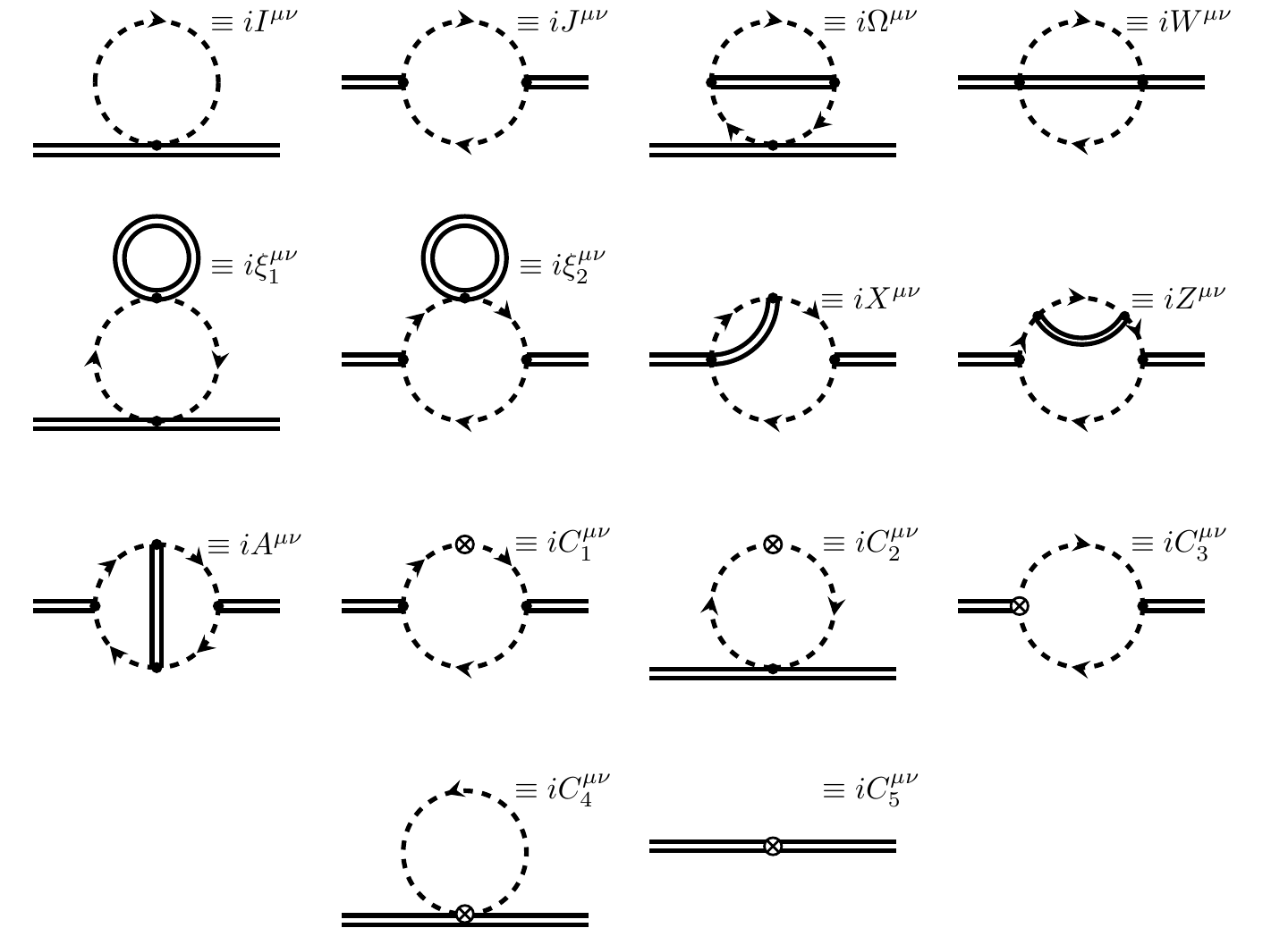}
	\caption{\label{2loopdiagrams} One-particle-irreducible diagrams contributing to the self-energy up to two-loop order.}
\end{figure*}

At two-loop order, i.e. $\cal{O}$$(g^4)$, the $\rho^0$ self-energy, denoted by $\Pi^{\mu\nu}$, is obtained by summing the diagrams shown in Figure \ref{2loopdiagrams} as follows
\eq{&\Pi\mn(p^2) = I\mn +J\mn +\frac{1}{2}\xi_1\mn +\xi_2\mn\\ &+\Omega\mn+2Z\mn+4X\mn +W\mn+A\mn\\
	&+2C_1\mn+C_2\mn+2C_3\mn+C_4\mn+C_5\mn\,\,, \this\label{sumd} 
} 
where $p$ is the external momentum to each diagram. Yet another consequence of the theory's Ward identities is that $\Pi^{\mu\nu}$ is transverse \cite{Hees,Mawande}
\eq{\Pi^{\mu\nu}&=\transv F_{\text{vac}}(p^2) \,\,. \this\label{selftrans} 
}
This means that $F_{\text{vac}}$ can be extracted by contracting $\Pi^{\mu\nu}$ with $g_{\mu\nu}$. Using Eq.\eqref{sumd} and Eq.\eqref{selftrans} gives
\eq{&F_{\text{vac}}(p^2)=\frac{1}{n-1}\left\lbrace g_{\mu\nu}I\mn+g_{\mu\nu}J\mn+\frac{1}{2}g_{\mu\nu}\xi_1\mn\right.\\ &
	+g_{\mu\nu}\xi_2\mn 
	+2g_{\mu\nu}Z\mn +4g_{\mu\nu}X\mn+g_{\mu\nu}W\mn
	\\
	&+g_{\mu\nu}A\mn+g_{\mu\nu}\Omega^{\mu\nu} +2g_{\mu\nu}C_1\mn +g_{\mu\nu}C_2\mn\\
	&+2g_{\mu\nu}C_3\mn+g_{\mu\nu}C_4\mn+g_{\mu\nu}C_5\mn\Big\rbrace\,\, . \this \label{projsum}
} 

It should be clear from Eq.\eqref{projsum} that the quantities of interest are the contractions of $g_{\mu\nu}$ with each of the diagrams. Before examining the integrals  arising from each of the Feynman diagrams we define
\eq{\alpha&= \frac{g^2}{16\pi^2} \,,\this \label{alpha}\\ 
	\mathscr{D}^nk&=\frac{d^nk}{i\pi^2\left(2\pi\mu\right)^{n-4}}\this\,\,.\label{measure}
}
It is to be understood that $k$ and $q$ represent loop momenta, while $p$ stands for the external momentum. 
The Feynman rules for the one-loop diagrams $I\mn$ and $J\mn$ yield
\eq{ I\mn &= 2 i g^2\mu^{2\epsilon}g\mn \int \frac{d^nk}{(2\pi)^n}\frac{1}{k^2-m^2}  \\
	&=-2\alpha g^{\mu\nu}\int \meask\frac{1}{k^2-m^2}\,\, , \this \label{I}
}
and 
\eq{ J^{\mu\nu} &= - i g^2\mu^{2\epsilon}\INTa\frac{(2k+p)^{\mu}(2k+p)^{\nu}}{\da\db}\\
	&=\alpha \int\meask \frac{(2k+p)^{\mu}(2k+p)^{\nu}}{\da\db}\, .\this \label{J}
}
Similarly, the two-loop diagrams involve 
\eq{\xi_1^{\mu\nu}&= 4n\alpha^2g^{\mu\nu} \intkq \frac{1}{\da^2(q^2-M^2)}\,\, ,\this\label{xi1}\\
&\\
\xi_2^{\mu\nu} &= 2n\alpha^2 \intkq\frac{(2k+p)^{\mu}(2k+p)^{\nu}}{D_1^2 D_2 D_4}\,\, ,\this\label{xi2} \\
	&\\
	\Omega^{\mu\nu}&=-2\alpha^2g^{\mu\nu} \intkq\frac{(k+q)^2}{{D_1} ^2{D_3} {D_4} }\,\, ,\this \label{omega}\\
	&\\
	Z^{\mu\nu}&=\alpha^2\intkq\frac{(k+q)^2(2k+p)^{\mu}(2k+p)^{\nu}}{{D_1} ^2{D_2} {D_3} {D_4} }\,\, ,\this \label{Z}\\
	&\\
	X^{\mu\nu}&=-2\alpha^2\intkq \frac{(k+q)^{\mu}(2k+p)^{\nu}}{{D_1} {D_2} {D_3} {D_4} }\,\, ,\this \label{X}\\
	&\\ 
	W^{\mu\nu}&=4\alpha^2g^{\mu\nu}\intkq\frac{1}{{D_2} {D_3} {D_4} }\,\, ,\this \label{W}
	}
	\eq{
A^{\mu\nu}&=\alpha^2\intkq \left\lbrace\frac{(k+q)\cdot(k+q+p) }{{D_1} {D_2} {D_3} {D_4} {D_5}}\right.\\
	&\times (2k+p)^{\mu}(2q+p)^{\nu} \bigg\}\,,\this \label{A}
} 
where 
\eq{D_1&= k^2-m^2\,\,, & D_2&=(k+p)^2-m^2\,, \\
	D_4&=q^2-m^2 \,\,,&D_3&=(k-q)^2-M^2\,\,,\\
	D_5&=(q+p)^2-m^2 \, .&&\this \label{denominators}
}
Finally, we consider diagrams with counter-term insertions. Applying the Feynman rules to $C_1^{\mu\nu}$ gives
 \eq{C_1\mn=\alpha\intkq\left\lbrace\frac{(2k+p)^{\mu}(2k+p)^{\nu}}{\da\db}\right. \\
 	\times\left(k^2\delta Z_2-m^2\delta Z_m\right)\bigg\rbrace\\
 	=\alpha\intkq\Bigg\lbrace\frac{(2k+p)^{\mu}(2k+p)^{\nu}}{\da\db} \\
 	\left.\times\left(k^2\sum_{j=1}^{\infty}\delta Z_2^{(j)}\alpha^j-m^2 \sum_{j=1}^{\infty}\delta Z_m^{(j)}\alpha^j\right)\right\rbrace \,\,,\this \label{c1series}
 }
where we invoked Eqs.\eqref{Z2series} and \eqref{Zmseries}. To order $\alpha^2$,
   each of the above series  is truncated at first order in $\alpha$, so that Eq.\eqref{c1series} becomes
 \eq{C_1\mn=\alpha^2\intkq\left\lbrace\frac{(2k+p)^{\mu}(2k+p)^{\nu}}{\da\db}\right. \\
 	\times\left(k^2\delta Z_2^{(1)}-m^2\delta Z_m^{(1)}\right)\bigg\rbrace\,\, . \this \label{C1}
 }
Proceeding similarly for the remaining  counter-term diagrams in Figure \ref{2loopdiagrams} gives
\eq{C_2\mn &= 2 \alpha^2 g\mn\intk \frac{k^2\delta Z_2^{(1)}-m^2\delta Z_m^{(1)}}{\da^2}\,\, ,\this \label{C2}\\
	&\\
	C_3\mn&= \alpha^2\delta Z_1^{(1)}\intk \frac{(2k+p)^{\mu}(2k+p)^{\nu}}{\da\db}\,\, ,\this \label{C3}\\
	&\\
	C_4\mn&= -2\alpha^2\delta {Z'_{1}}^{(1)}g\mn \intk \frac{1}{k^2-m^2}\,\, ,\this \label{C4}\\
	&\\
	C_5\mn&=\transv\left(-\alpha p^2\delta Z_{3}^{(1)}-\alpha^2 p^2\delta Z_3^{(2)} \right) \,.\this \label{C5} 
}
\subsection{Reduction of Feynman Integrals}
The strategy for evaluating $F_{\text{vac}}$ is to express each function in Eq.\eqref{projsum} in terms of well known scalar integrals. The one-loop integrals $g_{\mu\nu}I^{\mu\nu}$ and $g_{\mu\nu}J^{\mu\nu}$ are reduced, starting from Eqs.\eqref{I}-\eqref{J}, as follows
\eq{g_{\mu\nu}I^{\mu\nu}&=-2\alpha\int \meask\frac{g_{\mu\nu}g^{\mu\nu}}{k^2-m^2}\\
	&=-2n\alpha \int \meask\frac{1}{k^2-m^2}\\
	&=-2n\alpha A_{0}\left(m^2\right) \,\, ,\this \label{gI} 
} 
 and
 \eq{&g_{\mu\nu}J^{\mu\nu}= \alpha \int \meask\frac{\left(2k+p \right)^2}{\da\db}\\
 	&= \alpha \int \meask \frac{2D_1+2D_2+4m^2-p^2}{D_1D_2}\\
 	&=\alpha \left\lbrace 4A_0(m^2)+\left(4m^2-p^2 \right)B_0\left(p^2;m^2,m^2 \right) \right\rbrace,\this \label{gJ} 
 } 
 where we have introduced the one-loop basic integrals
 \eq{&A_0(m_i^2)=\int\meask\frac{1}{k^2-m_i^2} \, ,\this \label{A0int} \\
 	&B_0\left(p^2;m_i^2,m_j^2\right)=\int \frac{\meask}{(k^2-m_i^2)\left((k+p)^2-m_j^2\right)}\,.\this\label{B0int} 
 }
Both these integrals are known analytically \cite{passarino1979one}. Two-loop self-energy-type integrals may be similarly reduced in terms of a scalar basis, the so-called T-integrals \cite{weiglein1994reduction}. To define the T-integrals we first label the propagator momenta as
\eq{k_1&=k\,\,,& k_2&=k+p\,\,,&k_3&=k-q\,,\\
	k_4&=q\,\,,& k_5&=q+p\,\,.&&\this\label{propmom} 
}   
A general T-integral is then defined as 
\eq{&T_{i_1i_2\dots i_r}\left(p^2;m_{j_1}^2,m_{j_2}^2,\dots m_{j_r}^2 \right)\\
	&=\int\frac{\measkq}{(k_{i_1}^2-m_{j_1}^2)(k_{i_2}^2-m_{j_2}^2)\dots (k_{i_r}^2-m_{j_r}^2)}\,\,, \this\label{T-def}
}
where each $i_{l}\in \left\lbrace 1,2,3,4,5 \right\rbrace$ labels an internal momentum, and the $j_{z}$  are arbitrary indices labelling masses. For instance 
\eq{& T_{234}(p^2;m_a^2,m_b^2,m_c^2)\\&
	=\int  \frac{\measkq}{(k_2^2-m_a^2)(k_3^2-m_b^2)(k_4^2-m_c^2)}\\
	&=\int \frac{\measkq}{\left((k+p)^2-m_a^2 \right)\left((q-k)^2-m_b^2 \right)(q^2-m_c^2)}\,\this\label{50}.
}
Whenever the only factors entering the  denominator are those in Eq.\eqref{denominators}, we use the notation
\eq{T_{i_1i_2\dots i_r}=\int \measkq \frac{1}{D_{i_1}D_{i_2}\dots D_{i_r}} \,\,.\this\label{T-short} 
}
Additional integrals will be of the form
\eq{Y^{l_1\dots l_s}_{i_1\dots \i_r}&=\intkq \frac{k_{l_1}^2\dots k_{l_s}^2}{D_{i_1}\dots D_{i_r}} \,\,.\this\label{Y-def} 
}
These Y-integrals can be reduced to T-integrals as shown in \cite{weiglein1994reduction}. T-integrals are well-known, and there are various methods to evaluate them, both analytically and numerically \cite{kreimer1991master,davydychev1993two,scharf1994scalar,bauberger1995analytical}. 

We can now reduce the two-loop integrals  entering Eq.\eqref{projsum}.  From Eq.\eqref{W} we find
\eq{g_{\mu\nu}W^{\mu\nu}&= 4\alpha^2\intkq\frac{g_{\mu\nu}g^{\mu\nu}}{D_2D_3D_4}\\
	&=4n\alpha^2 T_{234} \,\, ,\this\label{gW} 
} 
and from Eq.\eqref{omega} 
\eq{&g_{\mu\nu}\Omega^{\mu\nu} =-2n\alpha^2\intkq\frac{(k+q)^2}{D_1^2D_3D_4}\\
	&=-2n\alpha^2\intkq \frac{2D_1+2D_4-D_3+4m^2-M^2}{D_1^2D_3D_4}\\
	&=n\alpha^2\left\lbrace 2T_{114}-4T_{134}-4T_{113} -2(4m^2-M^2)T_{1134} \right\rbrace.\this\label{gOmega}
}

Similarly, the remaining integrals in Eq.\eqref{projsum} are
\eq{&g_{\mu\nu}\xi_1^{\mu\nu}=4n^2\alpha^2T_{113}\,\, ,\this\label{gxi1}\\
&	\\
	&g_{\mu\nu}\xi_2^{\mu\nu}= -2n\alpha^2\left\lbrace (n-1)T_{123}+T_{113} \right\rbrace\,\,,\this\label{gxi2}\\
	&\\
	&g_{\mu\nu}X^{\mu\nu}=\alpha^2\left\lbrace -5T_{234}-T_{134}+2T_{124} -T_{123}\right.\\
&\left. -Y^1_{2345}-(7m^2-2M^2-2p^2)T_{1234}\right\rbrace\,\,,\this\label{gX}\\	
}
\eq{&g_{\mu\nu}Z^{\mu\nu}=\alpha^2 \left\lbrace 4T_{234}+4T_{134}-(n-1)T_{124}-T_{114}\right.\\
&+2(n-1)T_{123}+2T_{113}+2(8m^2-M^2-p^2)T_{1234}\\
&\left.+2(4m^2-M^2)T_{1134}+(4m^2-p^2)(4m^2-M^2)T_{11234}\right\rbrace, \this\label{gZ}
}
\eq{&g_{\mu\nu}A^{\mu\nu}=\alpha^2\left\lbrace 4Y^1_{2345}+4T_{234}+4T_{134}-8T_{124} +4T_{123} \right.\\
	&-(8m^2-2M^2-4p^2)T_{1245}+4(7m^2-3M^2-3p^2)\\
	&\left.\times T_{1234}+(4m^2-2M^2-p^2)(4m^2-M^2-2p^2)T_{12345} \right\rbrace\, .\this\label{gA} 
}
For the counter-term integrals we obtain
\eq{&g_{\mu\nu}C_1^{\mu\nu}=\\
	&-\alpha^2\left\lbrace C_{\pi}^{(1)}\left[(n-1)B_0(p^2;m^2,m^2)+\frac{n-2}{2m^2}A_0(m^2) \right] \right.\\
	&+\delta Z_2^{(1)}\left[4A_0(m^2)+(4m^2-p^2)B_0(p^2;m^2,m^2)\right]\bigg\rbrace\, ,\this \label{gC1} \\
	&\\
	&g_{\mu\nu}C_2^{\mu\nu}=2n\alpha^2\left(\delta Z_2^{(1)}+C_{\pi}^{(1)}\frac{n-2}{2m^2} \right)A_0(m^2)\,\,,\this\label{gC2} 
}
\eq{ g_{\mu\nu}C_3^{\mu\nu}&=\alpha^2\delta Z_{2}^{(1)}\left\lbrace (4m^2-p^2)B_0(p^2;m^2,m^2) \right. \\
	&\left.+4A_0(m^2)\right\rbrace \,\, ,\this\label{gC3}\\
	&\\
	g_{\mu\nu}C_4^{\mu\nu}&=-2n\alpha^2\delta Z_2^{(1)}A_0(m^2)\,\,,\this\label{gC4}\\
	&\\
	g_{\mu\nu}C_5^{\mu\nu}&=-(n-1)p^2\left\lbrace\alpha \delta Z_3^{(1)}+\alpha^2\delta Z_3^{(2)}\right\rbrace\,,\this\label{gC5} 
}
where 
\eq{C_{\pi}^{(1)}=m^2\left(\delta Z_2^{(1)}-\delta Z_m^{(1)} \right)\,\, .\this\label{cpi}
}
The combination of these counter-terms entering  Eq.\eqref{projsum} reduces to
\eq{&2g_{\mu\nu}C_1^{\mu\nu}+g_{\mu\nu}C_2^{\mu\nu}+2g_{\mu\nu}C_3^{\mu\nu}+g_{\mu\nu}C_4^{\mu\nu}+g_{\mu\nu}C_5^{\mu\nu}\\
	&=-\alpha(n-1)p^2\delta Z_{3}^{(1)}+\alpha^2(n-1)\bigg\lbrace -p^2\delta Z_3^{(2)} \\
	&\left.+2C_{\pi}^{(1)} \left[\frac{n-2}{2m^2}A_0(m^2)-B_0(p^2;m^2,m^2) \right]\right\rbrace \,\, . \this\label{counter-sum}
}  
\begin{figure}[b!]
	\centering
	\includegraphics{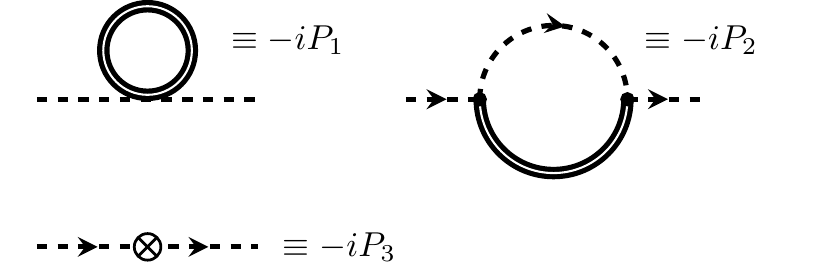} 
	\caption{\label{pi-self-diagrams} Diagrams contributing to the pion self-energy at one-loop order.} 
\end{figure}
To simplify Eq.\eqref{counter-sum} further, one requires expressions for the counter terms $\delta Z_2^{(1)}$ and $\delta Z_m^{(1)}$. These can be determined by evaluating the pion self-energy at the one-loop level. Interestingly, these counter terms appear in the form of $C_{\pi}^{(1)}$ in Eq.\eqref{counter-sum}, so that only their difference is required.  Hence, we consider the one-loop pion self-energy,  $\Pi_{\pi}(p^2)$  obtained by summing the diagrams in Figure \ref{pi-self-diagrams} 
\begin{equation}
-i\, \Pi_{\pi(1)} = -\frac{1}{2}\, i\, P_1 \,-i \,P_2 - i\, P_3\,, \label{Eq.65}
\end{equation}
where the symmetry factor  $\frac{1}{2}$ is associated with the neutral $\rho$-meson tadpole entering diagram $P_1$ in Fig.4. Application of the Feynman rules yields
\begin{equation}
-i \,P_1 = 2 \,i \,n \,\alpha \intk\dfrac{1}{(k^2-M^2)} \,,\label{Eq.66}
\end{equation}
\begin{equation}
-i \,P_2 = - i\, \alpha \,\intk \dfrac{ (k+2p)^2 }{ (k^2-m^2) [(k+p)^2-M^2]} \,,\label{Eq.67}
\end{equation}
\begin{equation}
-i \,P_3 = i \,\alpha ,\left(\p \delta Z_2^{(1)}-m^2\delta Z_m^{(1)} \right)\,, \label{Eq.68}
\end{equation}
which reduce to
\eq{P_1&=-2n\alpha A_0(M^2) \,\, ,\this \label{P1}\\
	P_2&=\alpha \left\lbrace (2\m-\M+2\p)B_0(\p;\M,\m) \right.\\
&\left. +2A_0(M^2)-A_0(m^2) 	\right\rbrace\,\, ,\this \label{P2}\\
	P_3&=-\alpha \left(\p\delta Z_2^{(1)}-\delta^{(1)}m^2 \right)\,\, ,\this \label{P3}
}	
giving
\eq{&\Pi_{\pi}(p^2)=\alpha\Big\lbrace(2m^2-M^2-2p^2)B_0(p^2;m^2,M^2) \\
	&\left. -(n-2)A_0(M^2)-A_0(m^2)-\left(\p\delta Z_2^{(1)}-m^2\delta Z_m^{(1)} \right)\right\rbrace \,. \this\label{pi-self-reduced}
}
Imposing the on-shell renormalization condition
\begin{equation}
\Pi_{\pi} \vert_{p^2=\mpi}=0 \,,
\end{equation}
implies that $m$, so far arbitrary,  is now the experimental  value of the pion mass \cite{PDG}, $m= 139.57\,\text{MeV}$, and   Eq.\eqref{pi-self-reduced} becomes
\begin{eqnarray}
    0 &=& \alpha \, \left[ -(n-2)A_0(M^2)-m^2\left(\delta Z_2^{(1)}-\delta Z_m^{(1)} \right) \right.\nonumber\\ [.3cm]
 &+&  \left.  (4m^2-M^2) \, B_0(m^2;M^2,m^2) - A_0(m^2) \right]\,,
 \label{Eq.76}
 \end{eqnarray}
implying
\begin{eqnarray}
C_\pi^{(1)}&=& (4m^2-\M)B_0(\m;\M,\m)\nonumber\\ [.3cm]
&-& A_0(m^2)     -(n-2)A_0(M^2)\,\, \label{cpi1}
\end{eqnarray}

Substituting Eq.\eqref{cpi1} into Eq.\eqref{counter-sum} gives
\eq{&2g_{\mu\nu}C_1^{\mu\nu}+g_{\mu\nu}C_2^{\mu\nu}+2g_{\mu\nu}C_3^{\mu\nu}+g_{\mu\nu}C_4^{\mu\nu}+g_{\mu\nu}C_5^{\mu\nu}\\
	&=-\alpha(n-1)p^2\delta Z_{3}^{(1)}+\alpha^2(n-1)\bigg\lbrace -p^2\delta Z_3^{(2)} \\
	&+2\left[(4m^2-\M)B_0(\m;\M,\m)-(n-2)A_0(M^2)\right.\\
	&\left.-A_0(m^2) \right]\times \left[\frac{n-2}{2m^2}A_0(m^2)-B_0(p^2;m^2,m^2) \right]\Big\rbrace\,.\this\label{counter-sum-reduced1}
} 
In  Eq.\eqref{counter-sum-reduced1} there are some products of one-loop basis integrals which can be expressed in terms of T-integrals to obtain
\eq{&2g_{\mu\nu}C_1^{\mu\nu}+g_{\mu\nu}C_2^{\mu\nu}+2g_{\mu\nu}C_3^{\mu\nu}+g_{\mu\nu}C_4^{\mu\nu}+g_{\mu\nu}C_5^{\mu\nu}\\
	&=-\alpha(n-1)p^2\delta Z_{3}^{(1)}+\alpha^2(n-1)\bigg\lbrace-2\left[T_{114}-T_{124}\right]\\
	&+2(4m^2-M^2)B_0(m^2;M^2,m^2)\left[ \frac{n-2}{2m^2}A_0(m^2)\right. \\ &-B_0(p^2;m^2,m^2) \bigg] -2(n-2)\left[T_{113}-T_{123}\right]\\
	&-p^2\delta Z_3^{(2)}\bigg\rbrace .\this\label{counter-sum-reduced2}
}  
We now consider the remaining contributions to $F_{\text{vac}}$ as expressed in Eq.\eqref{projsum}. Starting with the one-loop contributions, Eqs.\eqref{gI}-\eqref{gJ}, leads to
\eq{g_{\mu\nu}I^{\mu\nu}=\alpha\Big\lbrace 2(2-n)+(4m^2-p^2)B_0(p^2;m^2,m^2)\Big\rbrace,\this\label{1-loop-sum}
}  
and from  Eqs.\eqref{gxi1}-\eqref{gxi2} one finds
\eq{\frac{1}{2}g_{\mu\nu}\xi_1^{\mu\nu} +g_{\mu\nu}\xi_2^{\mu\nu}=2n(n-1)\alpha^2\left[T_{113}-T_{123}\right] \,.\this\label{1-loop-sep}
}
Using  Eqs.\eqref{gW}-\eqref{gOmega}, and Eqs.\eqref{gX}-  \eqref{gA} leads to 
 \eq{&2g_{\mu\nu}Z\mn +4g_{\mu\nu}X\mn+g_{\mu\nu}W\mn+g_{\mu\nu}A\mn +g_{\mu\nu}\Omega^{\mu\nu}\\
 	&=\alpha^2\Big\lbrace -4(n-1)\left[T_{113}-T_{123}\right] +2(n-1)\left[T_{114}-T_{124} \right] \\
 	&+4(n-2)\left[T_{234}-T_{134}\right]-2(n-2)(4m^2-M^2)T_{1134}\\
 	&+4(8m^2-2M^2-2p^2)T_{1234}-(8m^2-2M^2-4p^2)T_{1245}\\
 	&+(4m^2-2M^2-p^2)(4m^2-M^2-2p^2)T_{12345} \\
 	&+2(4m^2-p^2)(4m^2-M^2)T_{11234}\Big\rbrace\,\,.\this\label{pure2sum}
 }
 This completes all the information needed for $F_{\text{vac}}$,  Eq.\eqref{projsum}.  
 Using Eqs.\eqref{counter-sum-reduced2}-\eqref{pure2sum} gives
  \eq{&F_{\text{vac}}=\Big[ (4m^2-p^2)B_0(p^2;m^2,m^2)-2(n-2)A_0(m^2)\Big]\\
 	&\times\frac{\alpha}{n-1}-\alpha\, p^2\delta Z_3^{(1)} +\frac{\alpha^2}{n-1}\Big[4(n-2)T_{234}-4(n-2)T_{134}  \\
 	&-2(n-2)(4m^2-M^2)T_{1134}-(8m^2-2M^2-4p^2)T_{1245}\\
 	&+4(8m^2-2M^2-2p^2)T_{1234}+2(4m^2-p^2)(4m^2-M^2)\\
 	&\times T_{11234}+(4m^2-2M^2-p^2)(4m^2-M^2-2p^2)T_{12345}\Big]\\
 	&+\alpha^2T_{\otimes}
 	 -\alpha^2 p^2\delta Z_3^{(2)}\,\, ,\this\label{Fvac-reduced}
 } 
where 
\eq{T_{\otimes}&=2(4m^2-\M)\left(\frac{n-2}{2\m}A_0(\m)-B_0(\p;\m,\m) \right)\\
&\times B_0(\m;\M,\m)\,\, .\this \label{Txdef}
}
Equation Eq.\eqref{Fvac-reduced} gives the $\rho^0$ self-energy at  two-loop order in terms of known scalar integrals, to be evaluated next.
\subsection{Evaluation of the Basis Scalar Integrals}
The scalar integrals entering Eq.\eqref{Fvac-reduced} are well known, either analytically or numerically \cite{kreimer1991master,davydychev1993two,berends1994numerical,bauberger1995analytical}. Before writing their expressions,  the following definitions are needed
\eq{L_M&=\gamma_{\text{E}}+\ln\left( \frac{M^2}{4\pi\mu^2}\right)\,\, ,\this \\
	L_m&=\gamma_{\text{E}}+\ln\left( \frac{m^2}{4\pi\mu^2}\right)\,\, ,\this\\
	L_{|p|}&=\gamma_{\text{E}}+\ln\left( \frac{\left|p^2\right|}{4\pi\mu^2}\right)\,\, ,\this 
}
where $\gamma_{\text{E}}$ is Euler's constant. Next, we define
\eq{x&=\frac{p^2}{M^2}\,, \this \\
	y&=\frac{\m}{\M}\,\, ,\this\\
	R&=\frac{\m(r_1-r_2)}{p^2}\ln{r_1}\,\, ,\this \label{R}\\ 
	\tilde{R}&=\frac{1}{2}\frac{\tilde{r}_1-\tilde{r}_2}{x}\big( \ln{\tilde{r}_1}-\ln{\tilde{r}_2} \big)\,\, , \this \label{Rt}
} 
with
\eq{\ra&=\frac{1}{2}\left\lbrace 2-\frac{\p}{\m}+\sqrt{\left(2-\frac{\p}{\m} \right)^2-4}\right\rbrace \, \, ,\this\\
	\rb&= \frac{1}{2}\left\lbrace 2-\frac{\p}{\m}-\sqrt{\left(2-\frac{\p}{\m} \right)^2-4}\right\rbrace\,\, ,\this \\
	\rta&=\frac{1}{2}\big\lbrace 1+y-x+\sqrt{(x-y-1)^2-4y} \big\rbrace \, \, ,\this\\
	\rtb&=\frac{1}{2}\big\lbrace 1+y-x-\sqrt{(x-y-1)^2-4y} \big\rbrace  \,\, .\this
}
Finally,
\eq{w&=\frac{1-\rta}{\rtb-\rta} & \tilde{w}&=\frac{1-\rtb}{\rta-\rtb}\,\,\,\, ,\this \\
	z&=\frac{\rta(1-\rtb)}{\rta-\rtb} &\tilde{z}&=\frac{\rtb(1-\rta)}{\rtb-\rta}\,\,\,\, , \this \\
	w_0&=\frac{1-\ra}{\rb-\ra} &\tilde{w}_0&=\frac{1-\rb}{\ra-\rb}\,\,\,\, ,\this \\
	z_0&=\frac{\ra(1-\rb)}{\ra-\rb}& \tilde{z}_0&=\frac{\rb(1-\ra)}{\rb-\ra}\,\,\,\, . \this
}
The $\eta$-function, needed for the addition of logarithms with complex arguments, is given by
\eq{\eta(a,b)&= \ln(ab)-\ln a -\ln b\\
	&=2\pi i \big[\theta(-\text{Im}a )\theta(-\text{Im}b )\theta(\text{Im}(ab)) \\
	&-\theta(\text{Im}a )\theta(\text{Im}b )\theta(-\text{Im}(ab))  \big]\, .\this}
The expressions for the one-loop scalar integrals are \cite{passarino1979one,nierste1993two,scharf1994scalar,denner2007techniques}
\eq{A_0(\m)&=\intk \frac{1}{k^2-\m}\\
	&=\frac{\m}{\epsilon}+\m (1-L_{m})\\
	&+\epsilon\m \left\lbrace \frac{1}{2}\zeta(2)+\frac{1}{2}L_m^2-L_m+1 \right\rbrace \,\,, \this \label{A0Formula}
}
\eq{&B_0(\p;\m,\M)=\intk\frac{1}{\da((k+p)-\M)}\\
	&=\frac{1}{\epsilon}-\frac{1}{2}(L_m+L_M)+2-\frac{y-1}{2x}\ln y +\tilde{R}\\
	&+\frac{\epsilon}{2}\left\lbrace \zeta(2)+8+\frac{1}{4}(L_m+L_M)^2+\frac{1}{4}\ln^2y +4\tilde{R} \right.\\
	&+\big(L_m+L_M\big)\left(-2+\frac{y-1}{2x}\ln y-\tilde{R} \right) \\
	&-\frac{2(y-1)}{x}\ln y+\frac{\rta-\rtb}{x}\bigg[ \ln w\ln z-\ln \tilde{w}\ln \tilde{z}\\
	&+\li(z)-\li(\tilde{z})+\li(w)-\li(\tilde{w})\bigg]\bigg\rbrace\,\, ,\this 
}
\eq{&B_0(\p;\m,\m)=\frac{1}{\epsilon}+R+2-L_m+\frac{\epsilon}{2}\bigg\lbrace \zeta(2)\\
	&+8+L_m^2-2(R+2)+4R
	+\frac{\m(\ra-\rb)}{\p}\\
	&\times\bigg[ \ln w_0\ln z_0-\ln \tilde{w}_0\ln \tilde{z}_0+\li(z_0)-\li(\tilde{z}_0)\\
	&+\li(w_0)-\li(\tilde{w}_0) \bigg]\bigg\rbrace \,\, .\this \label{B0mmFormula}
}
The following  products of one-loop integrals enter in Eq.\eqref{Fvac-reduced}
	\eq{&T_{1245}=\left(B_0(\p;\m,\m)\right)^2\\
	&T_{\otimes}=2\left(\frac{n-2}{2\m}A_0(\m)-B_0(\p;\m,\m) \right)\\
	&\times(4m^2-\M) B_0(\m;\M,\m)\, ,\this
	}\label{Eq.(102)}
leading to 
	\eq{&T_{1245}= \frac{1}{\epsilon^2} +\frac{1}{\epsilon}\left(2R+4-2L_m\right)+2L_{m}^2 \\ &-4(R+2)L_m+R^2+8R+12+\zeta(2) \\
	&+\frac{m^2}{p^2}(r_1-r_2)\Big\lbrace \ln w_0\ln z_0-\ln \tilde{w}_0 \ln \tilde{z}_0\\ 
	&+\li (z_0)-\li (\tilde{z}_0)+\li(w_0)-\li(\tilde{w}_0) \Big\rbrace \, ,\this\\
	}
\eq{&T_{\otimes }=(4m^2-M^2)(R+2)\left\lbrace-\frac{2}{\epsilon}+3L_m+L_M\right\rbrace\\
& -\frac{m^2}{p^2}\left(4m^2-M^2 \right)(r_1-r_2) \bigg\lbrace \ln w_0\ln z_0 +\li (z_0)\\
&-\ln \tilde{w}_0 \ln \tilde{z}_0 -\li (\tilde{z}_0)+\li(w_0)-\li(\tilde{w}_0)\bigg\rbrace\\
&+\left\lbrace \left(1-\frac{M^2}{m^2}\right)\ln \left(\frac{m^2}{M^2}\right)-8-\left.\tilde{R}\right|_{p^2=m^2}\right\rbrace \\
\\
&\times \left(4m^2-M^2\right)(R+2)\, .\this}

The vacuum integrals $T_{134}$ and $T_{1134}$ are: (see \cite{van1984two,hoogeveen1985influence,ford1992standard,davydychev1993two,berends1994numerical})
\eq{&T_{134}=\frac{1}{2\epsilon^2}(2m^2+M^2)+\frac{1}{\epsilon}\left\lbrace \frac{3}{2}(2m^2+M^2)-2m^2L_m\right.\\
&-M^2L_M \Big\rbrace
+2m^2(L_m^2-3L_m)+M^2(L_M^2-3L_M)\\
&+\left(\frac{7}{2}+\frac{\zeta(2)}{2} \right)(2m^2+M^2)-\frac{M^2}{2}\ln ^2\left(\frac{m^2}{M^2}\right)\\
&+\frac{1}{2}\lambda \bigg\lbrace -4\li\left(\frac{1-\lambda}{2} \right) +2\ln ^2\left(\frac{1-\lambda}{2} \right)\\
&-\ln^2\left(\frac{m^2}{M^2}\right)+\frac{\pi^2}{3} \bigg\rbrace \,, \this \label{T134}
}
\eq{&T_{1134}=\frac{1}{2\epsilon^2} +\frac{1}{\epsilon}\left(\frac{1}{2}-L_m\right)+\frac{1}{2}+\frac{1}{2}\zeta(2)+L_m^2\\
&-L_m-\frac{1}{2\lambda} \bigg\lbrace -4\li\left(\frac{1-\lambda}{2} \right) +2\ln ^2\left(\frac{1-\lambda}{2} \right)\\
&-\ln^2\left(\frac{m^2}{M^2}\right)+\frac{\pi^2}{3} \bigg\rbrace\,\, , \this \label{T1134}
}  
where 
\eq{\lambda&=\sqrt{1-\frac{4m^2}{M^2}}\,\, . \this  
}
The master integral $T_{12345}$ is the only one that remains finite as $n\rightarrow 4$. It will be evaluated numerically using the following integral representation, first obtained in \cite{kreimer1991master}
\eq{&T_{12345}(p^2;m_{1}^2,m_2^2,m_3^2,m_4^2,m_5^2)\\
&= -\frac{1}{\pi^4}\int \bigg\{ \frac{d^4kd^4q}{(k^2-m_1^2)\left((k+p)^2-m_2^2 \right)\left((q-k)^2-m_3^2 \right)} \\ 
&\times \frac{1}{(q^2-m_4^2)\left((q+p)^2-m_5^2\right)}\bigg\}\\
&=-\frac{4}{p^2}\int\limits_{-\infty}^{\infty}dx \int\limits_{-\infty}^{\infty}dy\bigg\{\frac{1}{w_1^2-w_2^2} \frac{1}{w_4^2-w_5^2}\\
&\times\ln\left(\frac{(w_1+w_3+w_4)(w_2+w_3+w_5)}{(w_1+w_3+w_5)(w_2+w_3+w_4)} \right)\Bigg\}\,\, ,\this \label{T12345N}
} 
where
\eq{w_1&=\sqrt{x^2-\frac{m_1^2}{p^2}+i\varepsilon} \,\, , \this  \label{w1}\\
w_2&=\sqrt{(x+1)^2-\frac{m_2^2}{p^2}+i\varepsilon}\,\, ,\this \label{w2}\\
w_{3}&=\sqrt{(x+y)^2-\frac{m_3^2}{p^2}+i\varepsilon}\,\, ,\this \label{w3}\\
w_4&=\sqrt{y^2-\frac{m_4^2}{p^2}+i\varepsilon} \,\, , \this  \label{w4}\\
w_5&=\sqrt{(y-1)^2-\frac{m_5^2}{p^2}+i\varepsilon}\,\, .\this \label{w5}
}
Notice that this representation holds for arbitrary masses, so that to recover $T_{12345}$ in $n=4$ dimensions one sets $m_1=m_2=m_4=m_5=m$ and $m_3=M$.
\subsubsection*{The Semi-Numerical Algorithm}
The remainder of the $p^2$-dependent integrals, $T_{234}, T_{1234}, T_{11234},$ will be evaluated using the semi-numerical algorithm described in \cite{berends1994numerical}, in conjunction with some analytically calculated integrals from \cite{scharf1994scalar}. The goal of the algorithm is to write a $T$ integral  as a sum $T_{A}+T_N$, where $T_A$ involves massless propagators,  thus expressed analytically, and $T_N$ is finite in $n=4$ dimensions, thus evaluated numerically. We consider the case of  $T_{234}$ with arbitrary masses
\eq{ &T_{234}(p^2;m_2^2,m_3^2,m_4^2)\\\\
&=\int  \frac{\measkq}{\left((k+p)^2 -m_2^2\right)\left((k-q)^2-m_3^2\right)(q^2-m_4^2)}\,\, .\this \label{t234alg} 
} 
The algorithm entails a substitution using the following simple identities
\eq{ \frac{1}{(k-q)^2-m_3^2}&=\frac{1}{(k-q)^2} +\frac{m_3^2}{(k-q)^2\left((k-q)^2-m_3^2\right)}\,\,  \this \label{idty1}\\ \\
\frac{1}{q^2-m_4^2}&= \frac{1}{q^2}+\frac{m_4^2}{q^2(q^2-m_4^2)}\,\, .\this \label{idty2}
}
On the right-hand side of each of the above equations the first term replaces a massive propagator with a massless one and the second term decreases the degree of divergence of the integral. Equation \eqref{t234alg} then simplifies to
\eq{ &T_{234}(p^2;m_2^2,m_3^2,m_4^2)=T_{234}(p^2;m_2^2,m_3^2,0)\\
&+T_{234}(p^2;m_2^2,0,m_4^2)-T_{234}(m_2^2,0,0)\\
&+m_3^2m_4^2T_{23344}(p^2;m_2^2,m_3^2,0,m_4^2,0)\,\, .\this \label{t234breakdown}
}
The first three terms on the right-hand side of Eq.\eqref{t234breakdown} have analytic expressions to be found in the literature \cite{berends1994numerical,scharf1994scalar}, and the last term is finite for $n\rightarrow 4$. Hence, the goal of the algorithm is achieved in the case of $T_{234}$, leading to 
\eq{ &T_{234A}(p^2;m_2^2,m_3^2,m_4^2)=T_{234}(p^2;m_2^2,m_3^2,0)\\\\
&+T_{234}(p^2;m_2^2,0,m_4^2)-T_{234}(m_2^2,0,0) \,\, ,\this \label{t234A}
}
and
\eq{&T_{234N}(p^2;m_2^2,m_3^2,m_4^2)\\
&=m_3^2m_4^2T_{23344}(p^2;m_2^2,m_3^2,0,m_4^2,0)\, .\this\label{t234N} 
}
The remaining two integrals
\eq{&T_{1234}(p^2;m_1^2,m_2^2,m_3^2,m_4^2)\\
&=\int\measkq \bigg\{\frac{1}{(k^2-m_1^2)\left((k+p)^2-m_2^2\right)}\\
&\times \frac{1}{\left((k-q)^2-m_3^2\right)(q^2-m_4^2)}\bigg\}\,\, ,\this \\
\\
&T_{11234}(p^2;m_1^2,m_1^2,m_2^2,m_3^2,m_4^2)\\
&=\int\measkq \bigg\{\frac{1}{(k^2-m_1^2)^2\left((k+p)^2-m_2^2\right)}\\
&\times \frac{1}{\left((k-q)^2-m_3^2\right)(q^2-m_4^2)}\bigg\}\,\, ,\this
}
can be treated similarly using Eqs.\eqref{idty1}-\eqref{idty2} to give
\eq{T_{1234A}(p^2;m_1^2,m_2^2,m_3^2,m_4^2)&=T_{1234}(p^2;m_1^2,m_2^2,0,0)\,\, ,\this \label{t1234A}
}
\eq{ 
&T_{1234N}(p^2;m_1^2,m_2^2,m_3^2,m_4^2)\\
&= m_3^2T_{12334}(p^2;m_1^2,m_2^2,m_3^2,0,0)\\
&+m_4^2T_{12344}(p^2;m_1^2,m_2^2,0,m_4^2,0)\\
&+m_3^2m_4^2T_{123344}(p^2;m_1^2,m_2^2,m_3^2,0,m_4^2,0)\,\, , \this \label{t1234N}
}
as well as 
\eq{&T_{11234A}(p^2;m_1^2,m_1^2,m_2^2,m_3^2,m_4^2)\\
&=T_{11234}(p^2;m_1^2,m_1^2,m_2^2,0,0)\,\, , \this \label{t11234A} 
}
\eq{ 
&T_{11234N}(p^2;m_1^2,m_1^2,m_2^2,m_3^2,m_4^2)\\
&=m_3^2T_{112334}(p^2;m_1^2,m_1^2,m_2^2,m_3^2,0,0)\\
&+m_4^2T_{112344}(p^2;m_1^2,m_1^2,m_2^2,0,m_4^2,0)\\
&+m_3^2m_4^2T_{1123344}(p^2;m_1^2,m_1^2,m_2^2,m_3^2,0,m_4^2,0)\,\, . \this \label{t11234N}
}
Analytical expressions for Eq.\eqref{t1234A} and Eq.\eqref{t11234A} are derived in \cite{scharf1994scalar} using a combination of Cutkosky's cutting rules to extract the imaginary part of the integrals, and dispersion relations to recover their respective real parts. Equations \eqref{t1234N} and \eqref{t11234N} only involve finite integrals in the limit $n\rightarrow 4$ and so can be treated numerically. Berends and Tausk \cite{berends1994numerical} show  how to evaluate these integrals by adapting Kreimer's method \cite{,kreimer1991master} to obtain analogous double-integral representations 
\eq{&T_{234N}(p^2;m_2^2,m_3^2,m_4^2)=-4p^2\int\limits_{-\infty}^{\infty}dx\int\limits_{-\infty}^{\infty}dy\\
& \times \ln \left[\frac{(w_2+w_3+w_4)(w_2+\tilde{w}_3+\tilde{w}_4)}{(w_2+\tilde{w}_3+w_4)(w_2+w_3+\tilde{w}_4)}\right] \,\, ,\this \label{t234Nrep} 
}
\eq{ 
&T_{1234N}(p^2;m_1^2,m_2^2,m_3^2,m_4^2)=4 \int\limits_{-\infty}^{\infty}dx\int\limits_{-\infty}^{\infty}dy\\
&\frac{1}{w_1^2-w_2^2}\ln \left[\frac{(w_1+w_3+w_4)(w_2+\tilde{w}_3+\tilde{w}_4)}{(w_2+w_3+w_4)(w_1+\tilde{w}_3+\tilde{w}_4)}\right]\,\, ,\this \label{t1234Nrep}
}
\eq{
&T_{11234N}(p^2;m_1^2,m_1^2,m_2^2,m_3^2,m_4^2)=\frac{4}{p^2}\int\limits_{-\infty}^{\infty}dx\int\limits_{-\infty}^{\infty}dy\\
&\frac{1}{\left(w_1^2-w_2^2\right)^2}\left\lbrace\ln \left[\frac{(w_1+w_3+w_4)(w_2+\tilde{w}_3+\tilde{w}_4)}{(w_2+w_3+w_4)(w_1+\tilde{w}_3+\tilde{w}_4)}\right] \right.\\
&\left.-\frac{(w_1^2-w_2^2)(\tilde{w}_3+\tilde{w}_4-w_3-w_4)}{2w_1(w_1+w_3+w_4)(w_1+\tilde{w}_3+\tilde{w}_4)}\right\rbrace\,\, ,\this\label{t11234Nrep}
 }
where the new terms $\tilde{w}_3$ and $\tilde{w}_4$ are
 \eq{\tilde{w}_3&=\sqrt{(x+y)^2+i\epsilon}\,\, ,\this \label{wt3}\\ \\
 \tilde{w}_4&=\sqrt{y^2+i\epsilon}\,\, .\this \label{wt4}
 }
 The analytical parts of the $T$-integrals are given by \cite{berends1994numerical},\cite{scharf1994scalar}
 \eq{&T_{234A}=\frac{1}{2\epsilon^2}(2m^2+M^2)+\frac{1}{\epsilon}\left\lbrace 3m^2+\frac{3}{2}M^2\right.\\
 &\left.-2m^2L_m-M^2L_M-\frac{1}{4}p^2  \right\rbrace+M^2\left(L_M^2-3L_M\right) \\
 &+2m^2\left(L_m^2-3L_m \right) +\frac{1}{2}L_{|p|}+3(2m^2+M^2)\\
 &+\frac{1}{2}M^2\zeta(2)-\frac{1}{4}(m^2+M^2)\ln^2\left(\frac{m^2}{M^2}\right)\\
 &+\left\lbrace \li\left( \frac{m^2-M^2}{m^2}\right)-\li\left(\frac{M^2-m^2}{M^2} \right) \right\rbrace\\
 &\times\frac{1}{2}(m^2-M^2) -\frac{p^2}{4}\left\lbrace \ln\left|\frac{p^2}{m^2} \right|+\ln\left|\frac{p^2}{M^2}\right|+\frac{13}{2} \right\rbrace \\
 &+\frac{1}{4}p^2\left\lbrace \left(\frac{m^2}{p^2}\right)^2 -\left(\frac{M^2}{p^2}\right)^2\right\rbrace\ln\left(\frac{m^2}{M^2}\right)\\
 &+\frac{1}{2}m^2\left(\frac{m^2}{p^2}- \frac{p^2}{m^2}\right)\ln \left(1-\frac{p^2}{m^2}\right)-m^2\li\left(\frac{p^2}{m^2}\right) \\
 &-\frac{1}{2}(p^2+2m^2)R-\frac{1}{2}(p^2+m^2+M^2)\tilde{R}\\
 &+2m^2\left(1-\frac{m^2}{p^2}\right)\left\lbrace \li\left(1-r_1\right)+\li\left(1-r_2\right)\right\rbrace\\
 &+m^2\left(1-\frac{M^2}{p^2}\right)\bigg\lbrace \li\left(\frac{1-\tilde{r}_1}{-\tilde{r}_1}\right)+\li\left(\frac{1-\tilde{r}_2}{-\tilde{r}_2}\right)\\
 &-\li\left(\frac{m^2-M^2}{m^2}\right)\bigg\rbrace\\
 &+M^2\left(1-\frac{m^2}{p^2}\right)\bigg\lbrace\li(1-\tilde{r}_1)+\li(1-\tilde{r}_2)\\
 &-\li\left(\frac{M^2-m^2}{M^2}\right) \bigg\rbrace \,\ ,\this  \label{t234Aexpr}
 }.
 \eq{ 
 &T_{1234A}=\frac{1}{2\epsilon^2}+\frac{1}{\epsilon}\left\lbrace\frac{5}{2}-L_m+R \right\rbrace+\frac{19}{2}+\frac{3}{2}\zeta(2)\\
 &+L_m^2-(5+2R)L_m+\left(\frac{m^2}{p^2}-1 \right)\ln\left(1-\frac{p^2}{m^2}\right)\\
 &+\li\left(\frac{p^2}{m^2}\right)+4R+\frac{m^2(r_1-r_2)}{2p^2}\bigg\lbrace\ln^2(1+r_2)\\
 &-\ln^2(1+r_1)+2\li\left(\frac{1}{1+r_2}\right)\\
 &-2\li\left(\frac{1}{1+r_1}\right)-\li\left(1-r_1\right)+\li\left(1-r_2\right)  \\
 &-\li\left(r_2(1-r_2)\right)-\eta\left(1-\frac{p^2}{m^2},r_2\right)\ln\left(r_2(1-r_2)\right)\\
 &+\li\left(r_1(1-r_1)\right)+\eta\left(1-\frac{p^2}{m^2},r_1\right)\ln\left(r_1(1-r_1)\right)\bigg\rbrace ,\this \label{t1234Aexpr}
 }
 \eq{&T_{11234A}=\left\lbrace\frac{1}{\epsilon}-2L_m+\frac{p^2}{m^2}-2m^2 \right\rbrace \frac{R}{4m^2-p^2}\\
 &-\frac{1}{2p^2}\li\left(\frac{p^2}{m^2}\right)+\frac{1}{m^2}\left(\frac{m^2}{p^2}-1 \right)\ln\left(1-\frac{p^2}{m^2}\right)\\
 &+\frac{m^2(r_1-r_2)}{2p^2(4m^2-p^2)}\bigg\lbrace \ln^2(1+r_2)-\ln^2\left(1+r_1\right)\\
 &+2\li\left(\frac{1}{1+r_2}\right)-2\li\left(\frac{1}{1+r_2}\right)-\li(1-r_1)\\
  &+\li(1-r_2)
 -\li\left(r_2(1-r_2)\right) +\li\left(r_1(1-r_1)\right)\\
 &-\eta\left(1-\frac{p^2}{m^2},r_2\right)\ln\left(r_2(1-r_2)\right)\\
 &+\eta\left(1-\frac{p^2}{m^2},r_1\right)\ln\left(r_1(1-r_1)\right)\bigg\rbrace\,. \this \label{t11234Nexpr}
 }
 \subsection{Renormalization}
 We introduce the following subscript notation for the T-integrals
 \begin{equation}
T_{i_{1}\dots i_{j}}= T_{i_{1}\dots i_{j}D}+T_{i_{1}\dots i_{j}F}\,, 
 \end{equation}
 where $T_{i_{1}\dots i_{j}D}$ contains (i) the divergent part of $T_{i_{1}\dots i_{j}}$, and (ii) terms dependent on the renormalization scale.
  $T_{i_{1}\dots i_{j}F}$ contains the remainder of the finite (${\cal{O}}(\epsilon^0)$) terms of $T_{i_{1}\dots i_{j}}$. For some of the T-integrals, $T_{i_{1}\dots i_{j}F}$ will be at least partially evaluated numerically using Kreimer's double integral
 representation \cite{kreimer1991master}.  Turning to each of the integrals entering Eq.\eqref{Fvac-reduced}, we have
\eq{&T_{134D}=\frac{2m^2+M^2}{2\epsilon^2}+\frac{1}{\epsilon}\bigg\{
3m^2+\frac{3}{2}M^2 -2m^2L_m\\
&-M^2L_M \bigg\}+2m^2(L_m^2-3L_m)+M^2(L_M^2-3L_M)\, ,\this \label{T134D}\\
&T_{1134D}=\frac{1}{2\epsilon^2}+\frac{1}{\epsilon}\left\lbrace\frac{1}{2}-L_m\right\rbrace+L_m^2-L_m\, ,\this \label{T1134D}\\
&T_{1245D}= \frac{1}{\epsilon^2} +\frac{2}{\epsilon}\left(R+2-L_m\right)+2L_{m}^2-4(R+2)L_m \, , \this\label{T1245D}
}
\eq{T_{\otimes D}&=(4m^2-M^2)(R+2)\left\lbrace-\frac{2}{\epsilon}+3L_m+L_M\right\rbrace \,,\this\label{TxD}
}
\eq{ 
&T_{234D}= \frac{1}{2\epsilon^2}(2m^2+M^2) +\frac{1}{\epsilon}\bigg\lbrace 3m^2+\frac{3}{2}M^2\\
&-2m^2L_m-M^2L_M-\frac{1}{4}p^2 \bigg\rbrace+2m^2(L_m^2-3L_m)\\&+M^2(L_M^2-3L_M)+\frac{1}{2}p^2L_{\left|p\right|}\,\,\, ,\this\label{T234D}
}  
\eq{T_{1234D}&=\frac{1}{2\epsilon^2} +\frac{1}{\epsilon}\left\lbrace \frac{5}{2} - L_m+R \right\rbrace+L_m^2\\
&-(5+2R)L_m\, \, , \this \label{T1234D}\\ 
T_{11234D}&=\frac{R}{4m^2-p^2}\bigg\{\frac{1}{\epsilon}-2 L_m \bigg\} \,\, ,\this\label{T11234D}
} 
Returning to the rho-meson self-energy,  Eq.\eqref{Fvac-reduced} can be split into a one-loop and a two-loop contribution 
 \eq{F_{\text{vac}}&=\alpha F_{\text{vac}}^{(1)}+\alpha^2F_{\text{vac}}^{(2)}\,\, ,\this \label{FvacOrders}
     }
     with
     \eq{&F_{\text{vac}}^{(1)}= \frac{1}{3-2\epsilon}\Big\lbrace(4\m-\p)B_0(\p;\m,\m)\\
     &-4(1-\epsilon)A_0(\m) \Big\rbrace-\p\delta Z_3^{(1)}\,\, ,\this \label{Fvac1loop}\\ \\
     &F_{\text{vac}}^{(2)}=\frac{1}{3-2\epsilon}\Big\lbrace -8(1-\epsilon)T_{134}+8(1-\epsilon)T_{234}\\
     &-4(1-\epsilon)(4\m-\M)T_{1134}+4(8\m-2\M-2\p)\\
     & \times T_{1234}-(8\m-2\M-4\p)T_{1245}+2(4\m-\p)\\
     &\times(4\m-\M)T_{11234}+(4\m-2\M-\p)\\
     &\times(4\m-\M-2\p)T_{12345}\Big\rbrace
     +T_{\otimes}-\p\delta Z_3^{(2)}\,\, ,\this \label{Fvac2loop}
     }
     where $n=4-2\epsilon$ has been used, and $\alpha$ is defined in Eq.\eqref{alpha}.
     Imposing the on-shell renormalization condition
     \begin{equation}
     \text{Re} \left[ F_{\text vac}|_{p^2=M^2}\right] =0 \,\label{renormcondition0}   
     \end{equation}  
ensures that $M$ is the physical mass of the $\rho^0$-meson, $M= 775.5\, {\mbox{MeV}}$ \cite{PDG}, and   
     \eq{\left.\text{Re}\,F_{\text{vac}}^{(1)}\right|_{p^2=M^2}&=0\,\, ,\this \label{renormcondition1}\\
     \left.\text{Re}\,F_{\text{vac}}^{(2)}\right|_{p^2=M^2}&=0\,\, .\this \label{renormcondition2}
     }
     \subsubsection*{One-loop Contribution}
    The one-loop contribution to the self energy is 
     \eq{F_{\text{vac}}^{(1)}&=\frac{1}{3}\left(1+\frac{2\epsilon}{3}\right) \bigg\lbrace (4\m-\p) B_0(\p;\m,\m) \\
     &-4(1-\epsilon)A_0(\m) \bigg\rbrace -\p\delta Z_3^{(1)}+ O(\epsilon) \,\, . \this \label{Fvac1expanded}
     }
     Substituting this expression in Eqs.\eqref{A0Formula}, and \eqref{B0mmFormula}, $F_{\text{vac}}$ becomes
     \eq{F_{\text{vac}}^{(1)}&=\frac{1}{3}\left\lbrace -\frac{\p}{\epsilon}+\p L_m +8\m -\frac{8\p}{3}+(4\m-\p)R \right\rbrace \\
     &-\p\delta Z_3^{(1)}+O(\epsilon)\,\, .\this \label{Fvac1exp2}
     }
      Notice that $R$, Eq.\eqref{R}, can be written as
     \eq{&R=\left(1-\frac{4\m}{\p}\right)^{\frac{1}{2}} \Bigg\lbrace \ln \left|\frac{1-\sqrt{1-\frac{4\m}{\p}}}{1+\sqrt{1-\frac{4\m}{\p}} } \right|\\
     &+i\pi\theta\left( \p-4\m \right) \Bigg\rbrace
     +\left(\frac{4\m}{\p}-1\right)^{\frac{1}{2}}\cos^{-1}\left(1-\frac{\p}{2\m}\right)\\
     &\times\left[\theta(\p-4\m) -\theta(\p)\right]\,\, , \this \label{RReIm}
     }
     so that Eq.\eqref{Fvac1exp2} becomes
      \eq{&F_{\text{vac}}^{(1)}=\frac{1}{3}\Bigg\lbrace -\frac{\p}{\epsilon}+\p L_m +8\m -\frac{8\p}{3}\\& + \p \left(1-\frac{4\m}{\p}\right)^{\frac{3}{2}}\Bigg[\ln \left|\frac{1+\sqrt{1-\frac{4\m}{\p}}}{1-\sqrt{1-\frac{4\m}{\p}} } \right|\\
      &-i\pi\theta\left( \p-4\m \right) \Bigg] 
      +\p\left(\frac{4\m}{\p}-1\right)^{\frac{3}{2}}\cos^{-1}\left(1-\frac{\p}{2\m}\right)\\
      &\times\bigg[\theta(\p-4\m) -\theta(\p)\bigg]\Bigg\}-\p\delta Z_3^{(1)}+O(\epsilon)\,\, .\this
      \label{Eq.153}}
     The  on-shell renormalization condition, Eq.\eqref{renormcondition0}
      leads to  the one-loop contribution to the $\rho^0$ wave-function renormalization constant
     \eq{\delta Z_{3}^{(1)}&=\frac{1}{3} \Bigg\lbrace -\frac{1}{\epsilon} +L_m+\frac{8\m}{\M} -\frac{8}{3}+\left(1-\frac{4\m}{\M}\right)^{3/2}\\
     &\times\ln\left|\frac{1+\sqrt{1-\frac{4\m}{\M}}}{1-\sqrt{1-\frac{4\m}{\M}}}\right|\Bigg\rbrace +O(\epsilon)\,\, ,\this \label{dZ31loop}
     }
     as well as to the one loop contribution to the rho-meson self-energy in the limit $\epsilon\rightarrow 0$
      \eq{&F_{\text{vac}}^{(1)}=\frac{\p}{3}\Bigg\lbrace \left(1-\frac{4\m}{\p}\right)^{\frac{3}{2}} \Bigg[\ln \left|\frac{1+\sqrt{1-\frac{4\m}{\p}}}{1-\sqrt{1-\frac{4\m}{\p}} } \right|\\
      &-i\pi\theta\left( \p-4\m \right) \Bigg]+\frac{8\m}{\p}+\left(\frac{4\m}{\p}-1\right)^{\frac{3}{2}} \\
      & \times\cos^{-1}\left(1-\frac{\p}{2\m}\right)\bigg[\theta(\p-4\m) -\theta(\p)\bigg]+C\Bigg\rbrace\,\,  \this \label{Fvac1}
      }
      where
      \eq{C&=-\frac{8\m}{\M}+\left(1-\frac{4\m}{\M}\right)^{3/2}\ln\left|\frac{1-\sqrt{1-\frac{4\m}{\M}}}{1+\sqrt{1-\frac{4\m}{\M}}}\right|\,\, .\this  
      } 
      \subsubsection*{Two-loop Contributuion}
      Next, we consider the two-loop contribution to the self-energy. The cancellation of divergences is now less obvious. To analyse $F_{\text{vac}}^{(2)}$ we write it as 
      \eq{&F_{\text{vac}}^{(2)}=\frac{1}{3}\frac{1}{\left(1-\frac{2}{3}\epsilon \right)}\bigg\lbrace -8(1-\epsilon)(T_{134D}+T_{134F})\\
      &+8(1-\epsilon)(T_{234D}+T_{234F}) -4(1-\epsilon)(4m^2-M^2)\\
      &\times(T_{1134D}+T_{1134F}) +4(8m^2-2M^2-2p^2)\\&\times(T_{1234D}+T_{1234F})
       -(8m^2-2M^2-4p^2)\\
       &\times(T_{1245D}+T_{1245F})+2(4m^2-p^2)(4m^2-M^2)\\
       &\times(T_{11234D}+T_{11234F})+(4m^2-2M^2-p^2)\\
       &\times(4m^2-M^2-2p^2)T_{12345}\bigg\rbrace \\
      &+T_{\otimes D}+T_{\otimes F}-p^2 \delta Z_{3}^{(2)}= F^{(2)}_{D}+F^{(2)}_{F}-p^2 \delta Z_{3}^{(2)} \this \label{selfsplit} ,
      }
      where 
      \eq{&F^{(2)}_{D}=\frac{1}{3}\left(1+\frac{2}{3}\epsilon \right)\bigg\lbrace -8(1-\epsilon)T_{134D}\\
      &+8(1-\epsilon)T_{234D}-4(1-\epsilon)(4m^2-M^2)T_{1134D}\\
      & +4(8m^2-2M^2-2p^2)T_{1234D}-(8m^2-2M^2-4p^2)\\
      &\times T_{1245D}+2(4m^2-p^2)
      (4m^2-M^2)T_{11234D}\bigg\rbrace +T_{\otimes D}\,, \this \label{FD}}
      and
     \eq{ &F^{(2)}_{F}=\frac{1}{3}\bigg\lbrace-8T_{134F}+8T_{234F}-4(4m^2-M^2)T_{1134F}\\
      &+4(8m^2-2M^2-2p^2)T_{1234F}-(8m^2-2M^2-4p^2)\\
     & \times T_{1245F}
      +2(4m^2-p^2)(4m^2-M^2)T_{11234F}\\
      &+(4m^2-2M^2-p^2)(4m^2-M^2-2p^2)T_{12345}\bigg\rbrace +T_{\otimes F}\,\, . \this \label{Ffinite} 
      }
     Next, we consider the divergent terms in $F_{\text{vac}}^{(2)}$, i.e. only $F^{(2)}_{D}$ and $p^2 \delta Z_{3}^{(2)}$, as  $F^{(2)}_{F}$ is entirely finite.       
       Substituting Eqs. \eqref{T134D}-\eqref{T11234D} into Eq.\eqref{FD}, and after some  cancellations one finds
      \eq{&F^{(2)}_{D}=-\frac{2p^2}{\epsilon}+\frac{8p^2}{3} L_m +\frac{4p^2}{3} L_{|p|}\\
      &+\big( 4m^2-M^2\big)\left(\ln\left(\frac{M^2}{m^2}\right)+\frac{4}{3}\right)R\\
      &+2\big(4m^2-M^2 \big)\left(\ln\left(\frac{M^2}{m^2}\right)+\frac{5}{3}\right)-\frac{2p^2}{3}\this \,.
      }
      The divergent part of this result is cancelled by $p^2 \delta Z_{3}^{(2)}$, thus rendering the self energy finite. Notice that all terms proportional to $\dfrac{1}{\epsilon^2}$  vanish.
      Next, we determine $\delta Z_{3}^{(2)}$ explicitly using the renormalization condition Eq.\eqref{renormcondition2}. Equation \eqref{selfsplit} can be written as  
      \eq{&F_{\text{vac}}^{(2)}=-\frac{2p^2}{\epsilon}+\frac{8p^2}{3}L_{m}+\frac{4p^2}{3}L_{|p|}+\big(4m^2-M^2 \big)\\
      &
      \times \Bigg\lbrace\left(\ln\left(\frac{M^2}{m^2}\right)+\frac{4}{3}\right)R+2\left(\ln\left(\frac{M^2}{m^2}\right)+\frac{5}{3}\right)\Bigg\rbrace\\
      &-\frac{2p^2}{3}+F_{F}^{(2)}-p^2\delta Z_{3}^{(2)}\, .\this \label{F2} 
      }
    Imposing the on-mass-shell renormalization condition, Eq.\eqref{renormcondition0}, leads to
      \eq{ 
      & \delta Z_{3}^{(2)}=-\frac{2}{\epsilon}+\frac{8}{3}L_m+\frac{4}{3}L_M-\frac{2}{3} \\
      &+\left(\frac{4m^2}{M^2}-1\right)\Bigg\lbrace\left(\ln\left(\frac{M^2}{m^2}\right)+\frac{4}{3}\right)\text{Re}\left[\left. R\right|_{p^2=M^2}\right]  \\
      &+ 2\left(\ln\left(\frac{M^2}{m^2}\right)+\frac{5}{3}\right)\Bigg\rbrace+\frac{1}{M^2}\text{Re}\left[\left. F_{F}^{(2)}\right|_{p^2=M^2}\right]\,. \this \label{deltZ3}
      }
 Using     
      \eq{\text{Re} 
      \left[\left. R\right|_{p^2=M^2}\right]&=\sqrt{1-\frac{4m^2}{M^2}}\ln\left|\frac{1-\sqrt{1-\frac{4m^2}{M^2}}}{1+\sqrt{1-\frac{4m^2}{M^2}}} \right|\, ,\this \label{Eq.165} }
one obtains  
      \eq{\delta Z_{3}^{(2)}&=-\frac{2}{\epsilon}+\frac{8}{3}L_m+\frac{4}{3}L_M -\frac{2}{3}-\left(1-\frac{4m^2}{M^2}\right)^{3/2}\\
      &\times\left(\ln\left(\frac{M^2}{m^2}\right)+\frac{4}{3}\right)\ln\left|\frac{1-\sqrt{1-\frac{4m^2}{M^2}}}{1+\sqrt{1-\frac{4m^2}{M^2}}} \right|  \\
      &+ 2\left(\frac{4m^2}{M^2}-1\right)\left(\ln\left(\frac{M^2}{m^2}\right)+\frac{5}{3}\right)\\
      &+\frac{1}{M^2}\text{Re}\left[\left. F_{F}^{(2)}\right|_{p^2=M^2}\right]\, ,\this \label{dZ3}
      }
      where $F_{F}^{(2)}$ is defined in Eq.\eqref{Ffinite}. Substituting Eq.\eqref{dZ3}  into Eq.\eqref{F2} gives 
      \eq{F_{\text{vac}}^{(2)}&=f(p^2)+F_{F}^{(2)}\, ,\this \label{Fvac2}
      }
       where
      \eq{&f(p^2)=\frac{4p^2}{3}\ln\left|\frac{p^2}{M^2}\right| +\big(4m^2-M^2 \big)\left(\ln\left(\frac{M^2}{m^2}\right)+\frac{4}{3}\right)\\
      &\times\left\lbrace R-\frac{p^2}{M^2}\sqrt{1-\frac{4m^2}{M^2}}\ln\left|\frac{1-\sqrt{1-\frac{4m^2}{M^2}}}{1+\sqrt{1-\frac{4m^2}{M^2}}} \right|\right\rbrace \\
      &+2\big( 4m^2-M^2\big)\left(\ln\left(\frac{M^2}{m^2}\right)+\frac{5}{3}\right) \left(1-\frac{p^2}{M^2}\right)\\
      & -\frac{p^2}{M^2} \text{Re} \left[\left. F_{F}^{(2)}\right|_{p^2=M^2}\right] \, .\this \label{fp2}
      }
\section{Conclusions}       
In this paper we have extended the calculation of the vacuum polarization function of the KLZ theory from one-loop \cite{GK} to the two-loop level in perturbation theory. Applications of this result will be discussed separately.
    
\section{Acknowledgements}
This work was supported in part by the National Research Foundation (South Africa), the University of Cape Town, and the National Institute of Theoretical Physics (South Africa). Discussions with Hubert Spiesberger, Andreas von Manteuffel, and Gary Tupper are duly acknowledged. The authors thank Mikhail A. Ivanov for discussions, and for providing us his   numerical result of the finite master integral for comparison.


\begin{thebibliography}{99}
\bibitem{KLZ1} N.M. Kroll, T.D. Lee, B. Zumino, Phys. Rev. {\bf 175},  1376 (1967); J.H. Lowenstein, B. Schroer, Phys. Rev. D {\bf 6},  1553 (1972).
\bibitem{Hees} H. van Hees, hep-th/0305076 (unpublished); H. Ruegg, M. Ruiz-Altaba, Int. J. Mod. Phys. A {\bf 19}, 3265 (2004).
\bibitem{VMD} J.J. Sakurai, Ann. Phys. (N.Y.) {\bf 11}, 1 (1960); {\it ibid.} Currents and Mesons, University of Chicago Press (1969).
\bibitem{GK}  C. Gale, J. Kapusta, Nucl. Phys. B {\bf 357},  65 (1991).
\bibitem{GS}G. Gounaris, J.J. Sakurai, Phys. Rev. Lett. {\bf 21}, 244 (1968); see also M. Gourdin, Phys. Rep. {\bf 11 C}, 29 (1974).
\bibitem{tau} M. Davier, A. H\"{o}cker, Z. Zhang, Rev. Mod. Phys. {\bf 78}, 1043 (2006).
\bibitem{CAD1}C. A. Dominguez, J. I. Jottar, M. Loewe, and B. Willers, Phys. Rev. D {\bf 76}, 095002 (2007).
\bibitem{PDG} C. Patrignani {\it et al.}, Particle Data Group, Chin. Phys. C {\bf 40}, 100001 (2016).
\bibitem{rpiEXP} B. Ananthanarayan, I. Caprini, and D. Das, Phys. Rev. Lett. {\bf 119}, 132002 (2017).
\bibitem{CAD2}C. A. Dominguez, M. Loewe, and B. Willers, Phys. Rev. D {\bf 78}, 057901 (2008).
\bibitem{CAD3} C. A. Dominguez, M. Loewe, and M. Lushozi, Adv. High Energy Phys. {\bf 2015}, 803232 (2015).
\bibitem{LQCD1} S. Aoki, T. W. Chiu, H. Fukaya {\it et al.} Phys. Rev. D {\bf 80}, 034508 (2009).
\bibitem{LQCD2} V. G\"{u}lpers, G. von Hippel, and H. Wittig, Phys. Rev. D {\bf 89}, 094503 (2014).
 \bibitem{stueckelberg1938forces1}
E.C.G.~Stueckelberg,
Helv. Phys. Acta {\bf 11}, 225 (1938); Helv. Phys. Acta {\bf 11}, 299 (1938); Helv. Phys. Acta {\bf 11}, 312 (1938).
\bibitem{Mawande}
M.~Lushozi, PhD thesis, University of Cape Town, 2017.
\bibitem{passarino1979one} G. Passarino and M. Veltman, Nucl. Phys. B {\bf 160} 151 (1979).
\bibitem{weiglein1994reduction}
G.~Weiglein, R.~Scharf, and M.~B{\"o}hm, Nucl. Phys. B {\bf 416}, 606  (1994).
\bibitem{kreimer1991master}
D. Kreimer, Phys. Lett. B {\bf 273}, 277 (1991).
\bibitem{davydychev1993two}
A.I.~Davydychev and J.B.~Tausk, Nucl. Phys. B {\bf 397},123 (1993).
\bibitem{scharf1994scalar}
R.~Scharf and J.B.~Tausk, Nucl. Phys. B {\bf 412}, 523 (1994).
\bibitem{bauberger1995analytical}
S.~Bauberger, F.A.~Berends and M.~B{\"o}hm, M.~Buza,
 Nucl. Phys. B {\bf 434}, 383 (1995).
\bibitem{berends1994numerical}
F.A.~Berends and J.B.~Tausk, Nucl. Phys. B {\bf 421}, 456 (1994).
\bibitem{nierste1993two}
U.~Nierste, D.~M{\"u}ller, M.~B{\"o}hm.
Z.Phys. C {\bf 57}, 605 (1993).
\bibitem{denner2007techniques} A.~Denner,
Fortsch.Phys. {\bf 41}, 307 (1993). 
\bibitem{van1984two}
J.~Van~der Bij and M.~Veltman, Nucl. Phys. B {\bf 231}, 205 (1984).
\bibitem{hoogeveen1985influence}
F.~Hoogeveen, Nucl. Phys. B {\bf 259}, 19 (1985).
\bibitem{ford1992standard}
C.~Ford, I.~Jack, D.R.T.~Jones, Nucl. Phys. B {\bf 387}, 373 (1992).
\bibitem{besiii2016measurement}
BESIII collaboration et~al, Phys. Lett. B  {\bf 753}, 629  (2016).
\bibitem{LQCDg} C. Alexandrou {\it et al.}, Phys. Rev. D {\bf 96}, 034525 (2017).

\end{thebibliography}
\end{document}